\documentstyle[prb,aps,floats,epsfig]{revtex}%%prp%%
\begin{document}
\twocolumn[\hsize\textwidth\columnwidth\hsize\csname@twocolumnfalse%%prb%%
\endcsname%%prb%%
%----------------------------------------------------------
\title{Signatures of quantum integrability and nonintegrability \\ in the
spectral properties of finite Hamiltonian matrices}  

\author{Vyacheslav V. Stepanov and Gerhard M{\"u}ller} 

\address{Department of Physics, University of Rhode Island, Kingston RI
  02881-0817} 

\date{\today~--~3.5}
\maketitle

\begin{abstract}
  For a two-spin model which is (classically) integrable on a five-dimensional
  hypersurface in six-dimensional parameter space and for which level
  degeneracies occur exclusively (with one known exception) on four-dimensional
  manifolds embedded in the integrability hypersurface, we investigate the
  relations between symmetry, integrability, and the assignment of quantum
  numbers to eigenstates.  We calculate quantum invariants in the form of
  expectation values for selected operators and monitor their dependence on the
  Hamiltonian parameters along loops within, without, and across the
  integrability hypersurface in parameter space. We find clear-cut signatures of
  integrability and nonintegrability in the observed traces of quantum
  invariants evaluated in finite-dimensional invariant Hilbert subspaces. The
  results support the notion that quantum integrability depends on the existence
  of action operators as constituent elements of the Hamiltonian.
\end{abstract}
\twocolumn%%prb%%
%---------------------------------------------------------------
%end of one column-prb-style
]%%prb%%
%---------------------------------------------------------------
%gerhard-one-column-draft style, centered
%%goc%%\newpage
%%goc%%\narrowtext
%%goc%%\oddsidemargin=4cm
%%prp%%\onecolumn
%%%%%%%%%%%%%%%%%%%%%%%%%%%%%%%%%%%%%%%%%%%
%
\section{Introduction}\label{sec1}
%
%%%%%%%%%%%%%%%%%%%%%%%%%%%%%%%%%%%%%%%%%%%%
An autonomous classical Hamiltonian system with two degrees of freedom,
specified by some analytic function $H(p_{1},q_{1};p_{2},q_{2})$ of canonical
coordinates, is either integrable or nonintegrable -- tertium non datur. If a
second integral of the motion can be found, i.e. an analytic function
$I(p_{1},q_{1};p_{2},q_{2})$ which is functionally independent of $H$ and
satisfies $dI/dt=\{H,I\}=0$, the system is proven integrable. If chaotic
trajectories can be detected in the phase flow, the system is demonstrably
nonintegrable. Although it may happen that neither evidence can be ascertained
in practice for a given $H$, one or the other status is guaranteed to apply.

A question of long-standing interest has been whether an equally clear-cut
classification of systems exists in quantum mechanics.  Translating the
criterion of classical integrability into quantum mechanics for systems with few
degrees of freedom opens up loopholes of ambiguity that are not easily
closed.\cite{Weig92,WM95} Quantum mechanically, a second integral of the motion,
i.e. an operator $I$ with $[H,I]=0$ can always be constructed, for example, via
time average of an arbitrary operator $A$.\cite{Pere84,SM90} Performing the time
average in the energy representation eliminates all off-diagonal matrix elements
of $A$. Which attributes of quantum invariants are most sensitive to the
integrability status of the system?

Quantum chaos research has identified a catalog of attributes that distinguish
quantized nonintegrable from quantized integrable
systems.\cite{Gutz90,Reic92,Gutz98} The most widely studied distinctive
properties pertain to level statistics. However, in the extreme quantum limit of a typical
model system, where the density of energy levels is low, this distinction is
blurry at best or altogether unrecognizable. Only in the energy range where the
level density is high, which includes the semiclassical regime, do the
contrasting level spacing distributions come into focus.  Other indicators of
quantum chaos are similarly ambiguous.

One unequivocal discriminant between quantized integrable and nonintegrable
systems was recently identified in a study of level crossing manifolds in the
parameter space of a two-spin model.\cite{SM98} The system is specified by the
quadratic Hamiltonian
\begin{equation}\label{Ham}
H=\sum_{\alpha=x,y,z} \left\{ -J_{\alpha}S^{\alpha}_{1} S^{\alpha}_{2}+
\frac 12 A_{\alpha}
\left[ (S^{\alpha}_{1})^{2} +(S^{\alpha}_{2})^{2} \right] \right\}
\end{equation}
for two quantum spins ${\bf S}_1,{\bf S}_2$ of equal length
$\sqrt{\sigma(\sigma+1)}$ ($\sigma=\frac{1}{2},1,\frac{3}{2},\dots$). In the
classical limit $\hbar \to 0$, $\sigma \to \infty$,
$\hbar\sqrt{\sigma(\sigma+1)} = s$, the operators ${\bf S}_i$ turn into
3-component vectors, ${\bf S}_i = s(\sin\vartheta_i \cos\varphi_i$,
$\sin\vartheta_i \sin\varphi_i$, $\cos\vartheta_i )$, and Eq.~(\ref{Ham}) then
describes the energy function of an autonomous Hamiltonian system with two
degrees of freedom and canonical coordinates $p_i= s
\cos\vartheta_i,~~q_i=\varphi_i,~~i=1,2$. The classical integrability condition
was shown to have the form\cite{MTWKM87}
\begin{eqnarray}\label{Int}
 &&(A_{x}-A_{y})(A_{y}-A_{z})(A_{z}-A_{x}) \nonumber \\ 
&&\hspace*{2cm}+ \sum_{\alpha \beta \gamma = {\rm cycl}(xyz)} 
J^{2}_{\alpha}(A_{\beta}-A_{\gamma}) = 0.
\end{eqnarray}

Quantum mechanically, the Hamiltonian (\ref{Ham}) is expressible as a real
symmetric block-diagonal matrix, where each of the infinitely many
finite-dimensional blocks is associated with one spin-$\sigma$ realization of an
irreducible representation of the underlying (discrete) symmetry group (see
Appendix~\ref{appA}). 

The main conclusions of the level crossing study for this system may be
summarized as follows:\cite{SM98} (i) In the six-dimensional (6D) parameter
space of (\ref{Ham}), level degeneracies occur on smooth 4D structures.\cite{note4} (ii) For
an invariant block of $H$ with $K$ levels, this 4D structure consists of $K-1$
sheets, each representing one twofold $[k,k+1]$ level degeneracy in the sequence
$E_1\leq E_2\leq \ldots\leq E_K$. (iii) All 4D level crossing sheets are completely embedded
in the 5D integrability hypersurface. (iv) Under mild assumptions, the
integrability condition (\ref{Int}) can be determined analytically from the
conditions of level degeneracy in low-dimensional invariant Hilbert subspaces of
$H$.

These results strongly suggest that the notion of integrability remains
meaningful for quantum systems described by finite Hamiltonian matrices,
notwithstanding the fact that there exist universal algorithms for the
diagonalization of finite symmetric matrices.

For a deeper understanding of this subtle notion of quantum integrability, we
note that classical integrability guarantees the existence of a canonical
transformation $(p_1,q_1;p_2,q_2) \to (J_1,\theta_1;J_2,\theta_2)$ to
action-angle coordinates. It transforms the Hamiltonian $H(p_1,q_1;p_2,q_2)$ and
the second integral of the motion $I(p_1,q_1;p_2,q_2)$ into analytic functions
$H_C(J_{1},J_{2})$, $I_C(J_{1},J_{2})$. Each point $(J_{1}, J_{2})$ on the
action plane specifies a torus in phase space. In the nonintegrable
case, the actions $J_1,J_2$ are only defined for the surviving tori. Since the
tori are no longer dense anywhere in phase space, no smooth functions
$H_C,I_C$ on $J_1,J_2$ exist anymore. 

In a companion paper\cite{SM99} we have postulated that the underlying cause for
the embedment of $(d_I-1)$-dimensional level crossing manifolds in a
$d_I$-dimensional integrability manifold of the parameter space (with
dimensionality $d\geq d_I$) is linked to the existence of action operators as
constituent elements of the Hamiltonian. In that study we have demonstrated for
two distinct model systems the explicit functional dependence $H_Q(J_1,J_2)$,
$I_Q(J_1,J_2)$ of the Hamiltonian and the second integral of the motion on two
action operators, and compared it to the similar yet different functional
dependence $H_C(J_{1},J_{2})$, $I_C(J_{1},J_{2})$ of the corresponding classical
invariants on the classical action coordinates. A direct comparison was
facilitated by the fact that for the model systems considered there we knew not
only the second integral of the motion but also found a set of separable
canonical coordinates for the description of the classical time evolution.

%%%%%%%%%%%%%%%%%%%%%%%%%%%%%%%%%%%%%%%%%%%%
%
\section{Method}\label{sec2}
%
%%%%%%%%%%%%%%%%%%%%%%%%%%%%%%%%%%%%%%%%%%%%
A more indirect but no less compelling method for demonstrating the existence of
action operators as constituent elements of the quantum invariants $H,I$ in some
regions of parameter space, namely on the integrability hypersurface, and their
nonexistence elsewhere is pursued here for the two-spin model (\ref{Ham}). We
investigate the functional dependence of the eigenvalues of quantum invariants
on the Hamiltonian parameters, in particular across lines demarcating changes in
symmetry and/or integrability status.

On the integrability hypersurface (\ref{Int}), the natural quantum numbers of
the eigenstates within any invariant Hilbert subspace of $H$ are the integer
pairs $(m_{1},m_{2})$ specifying the eigenvalues (in units of $\hbar$) of the action
operators $J_1,J_2$.  Henceforth we call them {\it action quantum numbers}.
Elsewhere in parameter space, where level crossings between eigenstates of the
same parameter space are prohibited, the natural quantum number is a single
integer, the {\it energy sorting quantum number} $n$. What consequences do these
conflicting assignments of quantum numbers in the two regions of parameter space
have for the functional dependence of quantum invariants on the
Hamiltonian parameters?

Consider the case of a $K$-dimensional invariant subspace of (\ref{Ham}) spanned
by the basis given in Appendix~\ref{appA}. The $K$ eigenstates $|k\rangle, k=1,\ldots,K$
then form a star of orthonormal vectors pointing in oblique directions with
respect to the coordinate axes. A tiny change of the parameters $J_\alpha, A_\alpha$
causes the star of eigenvectors to rotate slightly. By monitoring the inner
product between eigenvectors before and after every infinitesimal parameter
change, we can keep track of all eigenvectors along the entire loop in parameter
space.

At the same time, we monitor the effect of the gradually transforming
eigenvectors on the eigenvalues of two quantum invariants. For this purpose we
choose the energy expectation value $E_k= \langle k|H|k\rangle$ and the expectation value
$I_k= \langle k|A|k\rangle$, where $A$ is some function of the $S_i^\alpha$.\cite{note3} When
the Hamiltonian parameters $J_\alpha, A_\alpha$ are varied along a path in 6D parameter
space, the vector $|k\rangle$ traces a path on the surface of a $K$-dimensional unit
sphere, and the point $(E_k, I_k)$ leaves a trace in the plane of invariants.

What if two eigenvectors are accidentally degenerate ($E_k =E_{k'}$), which
happens when their energy eigenvalues cross each other at some point on the path
in parameter space?  Generically, the eigenvalues of the second invariant
are different at the point of level degeneracy ($I_k \neq I_{k'}$). We can always
choose the second invariant such that this is the case.  At the crossing point
the orientation of the two eigenvectors is not fixed. However that ambiguity is
removed if we impose the condition that the path of every point $(E_k,I_k)$ in
the plane of invariants must be continuous.

We shall see that varying $J_\alpha, A_\alpha$ along a closed path in parameter space does
not guarantee that the trace of every eigenstate in the $(E_k,I_k)$-plane is
also closed. It may happen, for example, that two eigenvectors transform into
each other in the course of one parameter-space loop, thus leaving an open trace
in the plane of invariants, which will be closed only after a second traversal
of the loop. The two kinds of quantum numbers assigned to eigenstates in
different regions of parameter space as discussed previously, suggest the
following scenario.

(i) If the closed path in parameter space lies entirely on the integrability
hypersurface, then the traces of all eigenstates in the plane of invariants will
be closed. Along the loop, level crossings occur frequently, but the labeling of
all eigenstates by the action quantum numbers $m_1$, $m_2$ remains valid on every
stretch of it.

(ii) If the path in parameter space lies entirely off the integrability
hypersurface, the traces of all eigenstates will again be closed but for a
different reason. Level crossings are prohibited in this region. All states are
labeled by the energy sorting quantum number $n$. That label is valid along the
entire loop.

(iii) If the closed path in parameter space consists of a leg $A$ on and a leg
$B$ off the integrability hypersurface, then the conflicting assignment of
quantum numbers has the consequence that some of the traces in the plane of
invariants remain open. An eigenstate $\left|k \right>$ may undergo one or
several level crossings on leg $A$ of the path and thus end up at a different
position in the energy-level sequence at the beginning of leg $B$ when the
energy-sorting quantum number kicks in.  As the parameters are varied along leg
$B$ back to their starting values, the point $(E_k, I_k)$ is prevented from
finding its way back to the original position in the plane of invariants because
level crossings are now prohibited.

Not surprisingly, physical reality turns out to be more complicated. However,
the observations made by this method of analysis prove to be highly illuminating
in regard to the relations between symmetry, integrability, and the assignment
of quantum numbers.

%%%%%%%%%%%%%%%%%BEGIN-FIGURE%%%%%%%%%%%%%%%%%%%%%
\begin{figure}[t!]
\centerline{\hspace{3.5cm}\epsfig{file=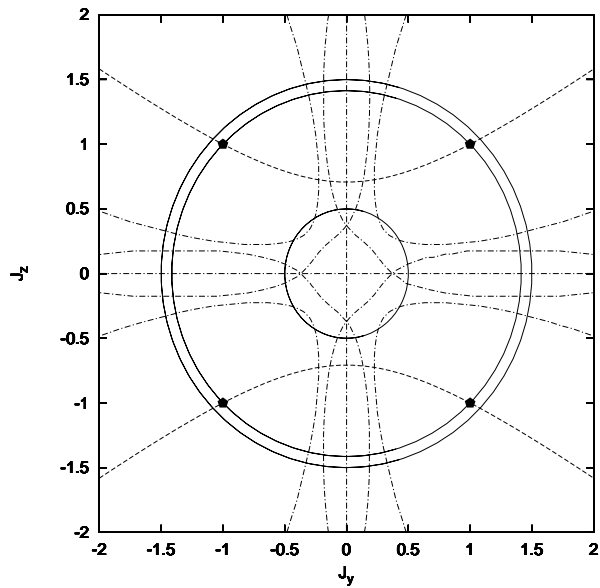,width=8.0cm,angle=-90}}
\caption{Reduced parameter space $(J_y,J_z,A)$ projected onto the
  $(J_y,J_z)$-plane. The two dashed lines mark the intersection,
  $2J_z^2-J_y^2=1$, of the integrability hyperboloid with the plane $A=0$. In
  the integrability plane $A=0$, level degeneracies of $H_{A1A}^5$ occur along
  the dot-dashed lines and multiple degeneracies at the symmetry points
  $|J_y|=|J_z|=1$ marked by the four pentagons. The solid circles represent
  projections of paths with radii $\sqrt{J_y^2+J_z^2}= \frac{1}{2}, \sqrt{2},
  \frac{3}{2}$ along which we track the quantum invariants $E_k, I_k$.}
\label{fig1}
\end{figure}
%%%%%%%%%%%%%%%%%%%%%END-FIGURE%%%%%%%%%%%%%%%%%%

%%%%%%%%%%%%%%%%%%%%%%%%%%%%%%%%%%%%%%%%%%%
%
\section{Results}\label{sec3}
%
%%%%%%%%%%%%%%%%%%%%%%%%%%%%%%%%%%%%%%%%%%%%
To facilitate comparison with results obtained previously, we use the same
reduced 3D parameter space as in Ref.~\onlinecite{SM98}. It is spanned by $J_y,
J_z, A_x-A_y \equiv 2A$ at $J_x=1, A_x+A_y=0, A_z=0$.\cite{note5} The integrability
condition (\ref{Int}), which becomes
\begin{equation}\label{Intred}
A(1 + J_y^2 - 2J_z^2 - 2A^2) = 0,
\end{equation}
is satisfied on a 2D surface consisting of the plane $A=0$ and a hyperboloid
with axis at $A=0, J_z=0$. Embedded in this integrability surface are 1D level
crossing manifolds in patterns whose complexity increases with the number of
levels in the invariant (Hilbert) subspaces under consideration.\cite{SM98}

Individual eigenstates $|k\rangle$ will now be tracked along closed paths in this
reduced parameter space. Each path selected displays distinct characteristic
features in the traces on the plane of invariants $(E_k,I_k)$. Here we use $I_k=
\langle k|(S_1^z+S_2^z)^2|k\rangle$.  We consider invariant (Hilbert) subspaces of symmetry
class A1A with $K=6,10$ levels corresponding to spin quantum numbers $\sigma=4,5$,
respectively (see Appendix \ref{appA}).

Figure~\ref{fig1} depicts the reduced parameter space projected onto the
integrability plane $A=0$. The dot-dashed lines represent the level crossing
manifold of $H_{A1A}^5$ with $K=10$ levels in the plane $A=0$.
None of the intersection points of two dot-dashed lines involves triple or
quadruple degeneracies. Each level crossing line can thus be labeled $[k,k+1]$
by the positions in the level sequence $E_1 \leq E_2 \leq \cdots \leq E_K$ of
the two levels involved in the crossing.\cite{note2}

%%%%%%%%%%%%%%%%%%%%BEGIN-FIGURE%%%%%%%%%%%%%%%%%%%
\begin{figure}[t!]
\centerline{\epsfig{file=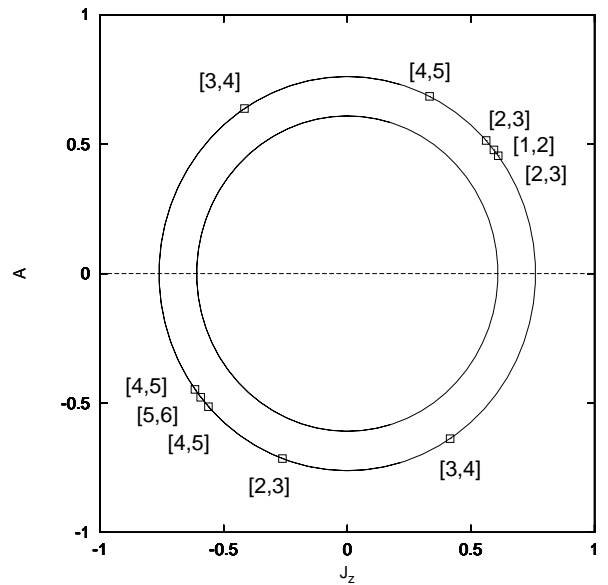,width=8.0cm,angle=-90}}
\caption{Reduced parameter space $(J_y,J_z,A)$ projected onto the
  $(J_z,A)$-plane. The solid circles represent paths with $J_z^2+A^2 = 0.3712,
  0.58$ at $J_y = 0.4$ along which we track the quantum invariants $E_k, I_k$.
  The larger circle is located on the integrability hyperboloid. The positions
  of level crossings of $H_{A1A}^4$ states along that path are indicated by
  squares. The dashed line marks the integrability plane $A=0$. }
\label{fig16}
\end{figure}
%%%%%%%%%%%%%%%%%%%%%END-FIGURE%%%%%%%%%%%%%%%%%%%

The integrability hyperboloid intersects the integrability plane along the two
dashed lines. There exist 30 $H_{A1A}^5$ level crossing lines on the
hyperboloid. These lines intersect the plane $A=0$ at seven points on each
dashed line, namely on the intersection points with dot-dashed lines
and on the symmetry points at $|J_y|=|J_z|=1$. The solid circles represent
projections of paths along which we track the quantum invariants $E_k,I_k$.

A different projection of the reduced parameter space is shown in Fig.
\ref{fig16}. The larger circle represents a path along the intersection of the
integrability hyperboloid with the plane $J_y=0.4$. The squares on that circle
mark the locations where the ten level crossing lines on the hyperboloid for
$H_{A1A}^4$ intersect the plane $J_y=0.4$. The smaller (concentric) circle
represents a path that is located in the nonintegrable region of parameter space
except for the two points where it intersects the integrability plane $A=0$
(dashed line).

%%%%%%%%%%%%%%%%%%%%%%%%%%%%%%%%%%%%%%%%%%%%%%
%
\subsection{Hallmark of integrability}\label{sec3A}
%
%%%%%%%%%%%%%%%%%%%%%%%%%%%%%%%%%%%%%%%%%%%%%%

%%%%%%%%%%%%%%%%%%%%BEGIN-FIGURE%%%%%%%%%%%%%%%%%%
\begin{figure}[t!]
\centerline{\epsfig{file=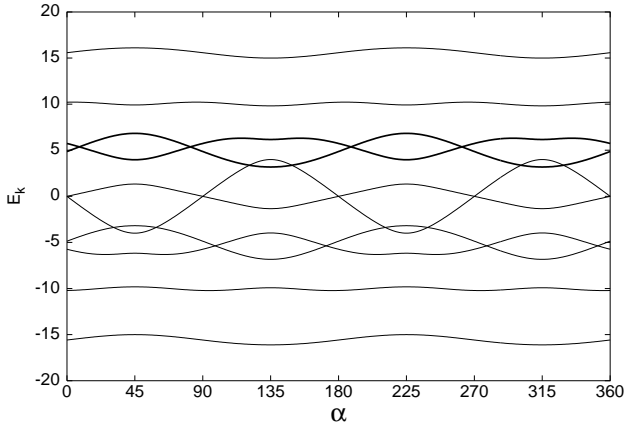,width=5.7cm,angle=-90}}
\caption{Energy eigenvalues $E_k, k=1,\ldots,10$ in the invariant subspace of
  $H_{A1A}^5$ as defined in Eq. (\ref{HRsigma}) and plotted versus the angular
  distance $\alpha$ on the circular path with radius $\sqrt{J_y^2+J_z^2}=0.5$ in the
  plane $A=0$ of the reduced parameter space $(J_y,J_z,A)$.}
\label{fig2}
\end{figure}
%%%%%%%%%%%%%%%%%%%%END-FIGURE%%%%%%%%%%%%%%%%%%

The first path considered is the circle $J_y^2+J_z^2=\frac{1}{4}$ in the plane
$A=0$ as shown in Fig.~\ref{fig1}. This path does not come close to any of the
symmetry points (pentagons). In Fig.~\ref{fig2} we have plotted the ten
levels of $H_{A1A}^5$ versus angular distance $\alpha$ on the circular path. We
observe 20 pairwise crossings between six levels at the angles where the
path intersects the dot-dashed lines in Fig.~\ref{fig1}.

No instances of level repulsion can be discerned in this plot, which is not to
say that the $\alpha$-dependence of adjacent levels is uncorrelated. Take the six
levels near the center of the spectrum. They can be divided into two groups of
three levels undergoing similar oscillations along the path. The synchronicity
of these oscillations is, in fact, a consequence of the (postulated) smooth
dependence of the functions $H_Q(J_1,J_2)$ and $I_Q(J_1,J_2)$ on $\alpha$ for this
path embedded in the integrability plane.\cite{SM99}

\begin{figure}[t!]
\centerline{\epsfig{file=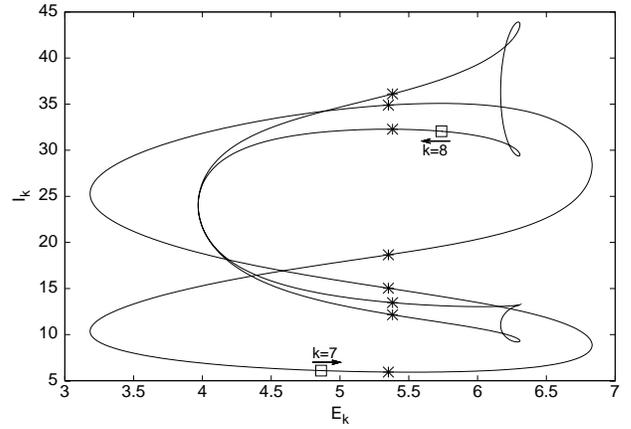,width=5.7cm,angle=-90}}
\caption{Closed traces in the $(E_k, I_k)$-plane of two $H_{A1A}^5$
  levels along the circular path with radius $\sqrt{J_y^2+J_z^2}=\frac{1}{2}$ in
  the plane $A=0$ of the reduced parameter space $(J_y,J_z,A)$. The traces start
  at the open squares $(\alpha=0^\circ)$ in the directions indicated. The asterisks mark
  level crossing points }
\label{fig3}
\end{figure}

In Fig.~\ref{fig3} we show the traces in the $(E_k, I_k)$-plane of the two
eigenstates whose levels undergo four [7,8] crossings along the path (thick
lines in Fig.~\ref{fig2}). The traces are continuous, closed, and smooth. The
square and the arrow indicate the starting point and the direction of the trace.
Every level crossing is represented by two vertically displaced asterisks, one
on each trace.

It is important to note that the traces remain perfectly smooth at the points of
level crossing. The level crossings have no impact on the eigenvectors, or on
the expectation values $I_k$. Every eigenvector loops around and returns to its
original orientation in Hilbert space. Its path is largely unaffected by the presence of
other eigenvectors which become instantaneously degenerate with it. It is as if
vectors undergoing level crossings belonged to different invariant subspaces.

%%%%%%%%%%%%%%%%%%%%BEGIN-FIGURE%%%%%%%%%%%%%%%%%
%%%%%%%%%%%%%%%%%%%%%END-FIGURE%%%%%%%%%%%%%%%%%

The behavior of energy levels as observed in Fig.~\ref{fig2} and the properties
of traces as seen in Fig.~\ref{fig3} reflect what we expect for a typical
situation in an integrable system with two degrees of freedom. The two
invariants $E_k, I_k$ are functions of two quantized actions $J_1,J_2$ with a
smooth dependence on the Hamiltonian parameters. The discrete values of the
actions define the natural quantum numbers of all levels, and each eigenstate
maintains its identity along any path in parameter space notwithstanding the
presence of level crossings. All traces produced along closed paths are
therefore closed as well.

There are two sources of complication forcing on us a refinement of this
description without undermining the postulated link between quantum integrability
and action operators. These two complications will be discussed next before we
investigate the effects of nonintegrability.

%%%%%%%%%%%%%%%%%%%%%%%%%%%%%%%%%%%%%%%%%%%%%
%
\subsection{Level repulsion near symmetry points}\label{sec3B}
%
%%%%%%%%%%%%%%%%%%%%%%%%%%%%%%%%%%%%%%%%%%%%%
The second path considered is the circle $J_y^2+J_z^2=\frac{9}{4}$ in the
integrability plane $A=0$ (see Fig.~\ref{fig1}). What makes it different from
the previous path is that it passes close to the four points $|J_y|=|J_z|=1$,
where additional degeneracies occur, caused by a higher symmetry.

\begin{figure}[t!]
\centerline{\epsfig{file=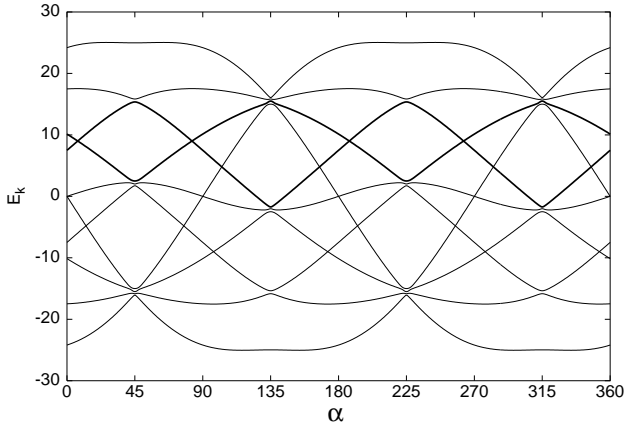,width=5.7cm,angle=-90}}
\caption{Energy eigenvalues $E_k, k=1,\ldots,10$ in the invariant subspace of
  $H_{A1A}^5$ plotted versus $\alpha$ on the path with radius
  $\sqrt{J_y^2+J_z^2}=1.5$ in the plane $A=0$ of $(J_y,J_z,A)$ space.}
\label{fig4}
\end{figure}

The ten levels of $H_{A1A}^5$ versus $\alpha$ are plotted in Fig.~\ref{fig4}. As
in Fig.~\ref{fig2} for the previous path, we observe 20 level crossings, each
one associated with a point where the circular path intersects one of the
dot-dashed lines in Fig.~\ref{fig1}. In addition to these crossings we observe
instances of level collisions at $\alpha = n\pi/2, n=1,3,5,7$, i.e. in the
vicinity of the symmetry points.

%%%%%%%%%%%%%%%%%%BEGIN-FIGURE%%%%%%%%%%%%%%%%%%%
%%%%%%%%%%%%%%%%%%%%%END-FIGURE%%%%%%%%%%%%%%%%%%

\begin{figure}[t!]
\centerline{\epsfig{file=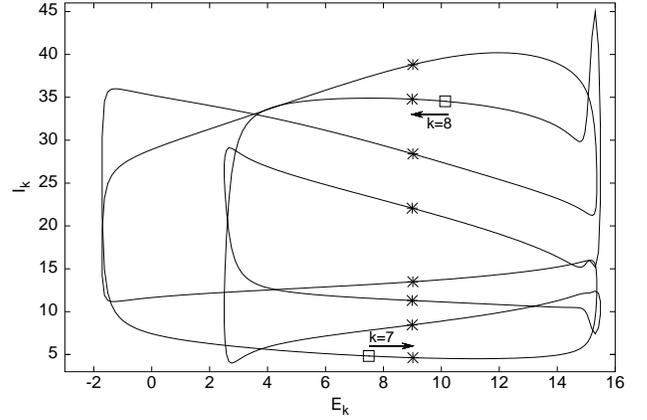,width=5.7cm,angle=-90}}
\caption{Closed traces in the $(E_k, I_k)$-plane of two $H_{A1A}^5$
  levels along the path with radius $\sqrt{J_y^2+J_z^2}=1.5$ in the plane $A=0$
  of $(J_y,J_z,A)$ space. The traces start at the squares $(\alpha=0^\circ)$ in the
  directions indicated. The asterisks mark level crossing points.}
\label{fig5}
\end{figure}

It is instructive to compare the effects of level crossings and level collisions
on the traces in the plane of invariants. In Fig.~\ref{fig5} we show again the
trace of the point $(E_k, I_k)$ for two states that are involved in four [7,8]
levels crossings (thick lines in Fig.~\ref{fig4}), now along the second path.
These traces exhibit features not seen in Fig.~\ref{fig3}.

We again observe that none of the level crossings leaves any mark on the traces,
implying that the wave functions of the two eigenstates are completely
unperturbed by the instantaneous level degeneracies (see asterisks). On any
stretch between successive mutual crossings, both levels collide with one
neighboring level, and each collision does have a dramatic effect on the traces
of the states involved in the collision.  Level collisions produce precipitous
changes in the second invariant $I_k$ near the closest encounter of the
colliding levels.  The rapid variation of expectation values signals a strong
perturbation of the wave functions in a level collision.  The presence of this
characteristic signature of level collisions is as conspicuous in the traces
shown in Fig. \ref{fig5} as is their absence in the traces shown in Fig.
\ref{fig3}.
%%%%%%%%%%%%%%%%%%%%%BEGIN-FIGURE%%%%%%%%%%%%%%%%%
%%%%%%%%%%%%%%%%%%%%%END-FIGURE%%%%%%%%%%%%%%%%%%

In what might be called a hard level collision, the two states exchange wave
functions in a manner like two billiard balls exchange momenta in a head-on
collision. This makes it hard to distinguish a hard collision from a crossing in
a plot such as Fig.~\ref{fig4} because of graphical resolution. A plot of one
invariant versus the other (Fig.~\ref{fig5}) is much more sensitive to that
distinction. Here a hard level collision produces a variation in $I_k$ that
looks almost like a discontinuity.

The phenomena observed in Figs. \ref{fig4} and \ref{fig5} are not in
contradiction with the assertion that the invariants $E_k, I_k$ are functions of
two quantum actions. It tells us, however, that the dependence of these
functions on the Hamiltonian parameters is singular at the symmetry points of
$H$. The phenomenon of level repulsion in the immediate vicinity of symmetry
points is then caused by invariants pertaining to the higher symmetry and by the
associated additional level degeneracies.

The traces of all levels depicted in Fig. \ref{fig4} are closed as were all
traces of the levels shown in Fig. \ref{fig2}. The implication is that the
number of crossings between any pair of levels must be an even number. The fact
is that neither the level crossings nor the level collisions can cause any
confusion in the labeling of the levels by action quantum numbers along a path
in the integrability plane $A=0$ as long as it avoids the points $|J_y|=|J_z|=1$
of higher symmetry with symmetry induced level degeneracies. Each eigenstate
maintains its identity along such paths, or so it seems.

%%%%%%%%%%%%%%%%%%%%%%%%%%%%%%%%%%%%%%%%%%%%%%
%
\subsection{Open traces caused by a change in symmetry}\label{sec3C}
\begin{figure}[t!]
\centerline{\epsfig{file=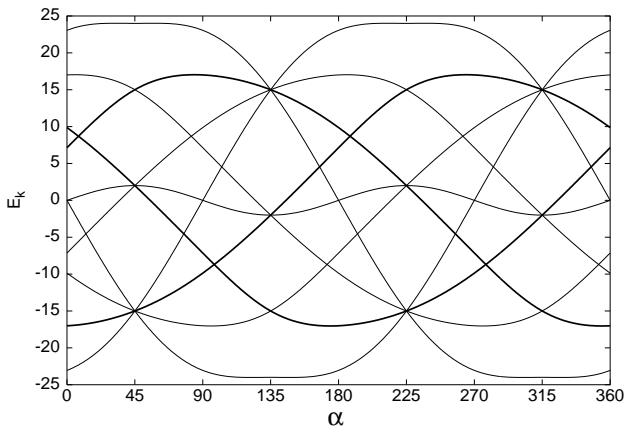,width=5.7cm,angle=-90}}
\caption{Energy eigenvalues $E_k, k,\ldots,10$ in the invariant subspace of
  $H_{A1A}^5$ plotted versus $\alpha$ on the path with radius
  $\sqrt{J_y^2+J_z^2}=\sqrt{2}$ in the plane $A=0$ of $(J_y,J_z,A)$ space.}
\label{fig6}
\end{figure}

%
%%%%%%%%%%%%%%%%%%%%%%%%%%%%%%%%%%%%%%%%%%%%%%
The third path considered is the circle $J_y^2+J_z^2=2$ at $A=0$ (see Fig.
\ref{fig1}). It is embedded in the integrability plane and passes through the
points $|J_y|=|J_z|=1$. The impact of these symmetry points on the energy levels
is depicted in Fig. \ref{fig6}. What were level collisions in Fig. \ref{fig4}
have now turned into additional level crossings. At the symmetry points, the ten
levels combine into a singlet, a doublet, a triplet, and a quadruplet. No
instances of level repulsion are observable anymore.

%%%%%%%%%%%%%%%%%%%BEGIN-FIGURE%%%%%%%%%%%%%%%%%%%%
%%%%%%%%%%%%%%%%%%%%END-FIGURE%%%%%%%%%%%%%%%%%%%

\begin{figure}[htb]
\centerline{\epsfig{file=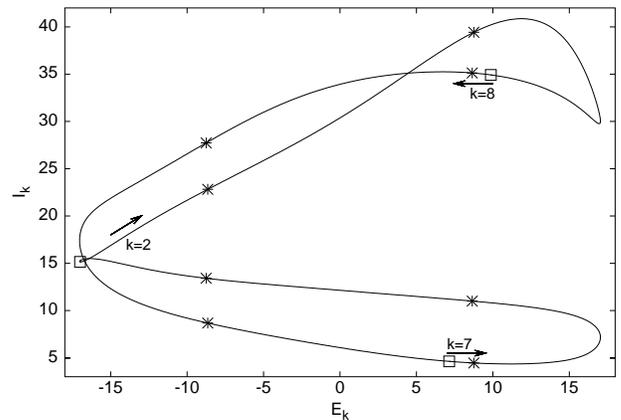,width=5.7cm,angle=-90}}
\caption{Open traces in the $(E_k, I_k)$-plane of three $H_{A1A}^5$
  levels along the path with radius $\sqrt{J_y^2+J_z^2}=\sqrt{2}$ in the plane
  $A=0$ of $(J_y,J_z,A)$ space. The traces start at the squares $(\alpha=0^\circ)$ in the
  directions indicated The asterisks mark level crossing points.}
\label{fig7}
\end{figure}

The absence of level collisions along this path is confirmed by a study of the
traces in the $(E_k, I_k)$-plane. In Fig. \ref{fig7} we show the traces of the
two states that again start in the seventh and eighth positions of the level
sequence. Gone are the rapid near-vertical displacements which we have
identified in Fig.  \ref{fig5} and which were caused by level collisions. The
traces in Fig. \ref{fig7} are as unaffected by the new symmetry-induced level
crossings as they are oblivious of crossings elsewhere in the integrability
plane.

However, a striking new feature makes its appearance in Fig. \ref{fig7}. The
traces do not close in themselves after one loop around the circular path in
parameter space. The eighth level becomes the seventh level after one loop, and
then turns into the second level after two loops. Only after the third loop does
it end up in the original eighth position of the level sequence.

In Fig. \ref{fig6} the three levels involved in that loop are drawn as thick
lines. Inspection shows that there are two further groups of three states which
transform into each other as the parameter values loop around the circle. That
leaves one state (near the center of the spectrum) whose trace closes in itself
after one loop.

Is this phenomenon of levels transforming into each other compatible with the
notion that the invariants are functions of the quantized actions with a smooth
dependence on the Hamiltonian parameters? Yes if we allow the
dependence on the parameters to be singular at points of higher symmetry within
the integrability manifold. The presence of such singularities was already
suggested by the level collisions observed in Figs. \ref{fig4} and \ref{fig5}.
The results of Figs. \ref{fig6} and \ref{fig7} confirm the singular parameter
dependence from a different vantage point.

When we start with the second path in parameter space (Sec. \ref{sec3B}) and
increase the radius of the circle gradually toward that of the third path, we
observe a gradual hardening of the level collisions near the symmetry points.
The hardening is characterized by increasingly sharp curvatures in the graphs of
$E_k$ versus $\alpha$ (Fig. \ref{fig4}) and by increasingly rapid vertical
variations in the graphs $I_k$ versus $E_k$ (Fig. \ref{fig5}). 
%%%%%%%%%%%%%%%%%%%BEGIN-FIGURE%%%%%%%%%%%%%%%%%%%
%%%%%%%%%%%%%%%%%%%%END-FIGURE%%%%%%%%%%%%%%%%%%%

In the limiting case of this path, the sharply curved but smooth bends in the
graph $E_k$ versus $\alpha$ turn into cusps, and the fast but smooth vertical
variations in the graphs $I_k$ versus $E_k$ turn into discontinuities.  An
infinitely hard level collision is indistinguishable from a level crossing. In
Figs. \ref{fig6} and \ref{fig7} smooth segments of graphs between singularities
that belong to different {\it colliding} levels are rejoined to form entirely
smooth graphs of {\it crossing} levels.

Hence, if we insist that all levels maintain their identity along any closed
path in the integrability plane $A=0$, we must interpret all level crossings
that take place at the points of higher symmetry, $|J_y|=|J_z|=1$, as infinitely
hard level collisions.  All the evidence accumulated thus far still supports the
existence of the functions $H_Q(J_1,J_2)$ and $I_Q(J_1,J_2)$ with a smooth
parameter dependence on the integrability manifold, provided we allow for
singularities at points of higher symmetry.

Before we discuss the strongly contrasting properties of quantum invariants
along paths that are not fully embedded in the integrability manifold of
(\ref{Ham}), we should report on yet another feature that complicates the
interpretation of the integrable cases.

%%%%%%%%%%%%%%%%%%%%%%%%%%%%%%%%%%%%%%%%%%%%%
%
\subsection{Open traces caused by topology}\label{sec3D}
%
%%%%%%%%%%%%%%%%%%%%%%%%%%%%%%%%%%%%%%%%%%%%%
The circle $A^2+J_z^2=0.58$ with center at $J_y=0.4$ is the fourth path along
which we study the behavior of quantum invariants. This path represents a
circular section of the integrability hyperboloid (\ref{Intred}) (see Fig.
\ref{fig16}). Like the first path considered, it does not pass near any point in
parameter space where symmetry induced level degeneracies occur.

\begin{figure}[htb]
\centerline{\epsfig{file=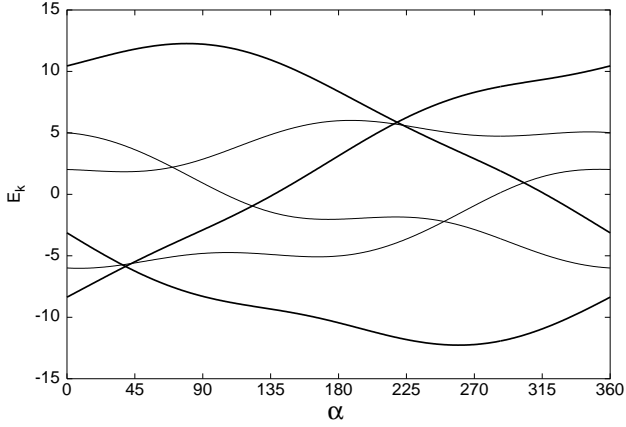,width=5.7cm,angle=-90}}
\caption{Energy eigenvalues $E_k, k=1,\ldots,6$ in the invariant subspace of
  $H_{A1A}^4$ plotted versus $\alpha$ on the path with radius
  $\sqrt{A^2+J_z^2}=\sqrt{0.58}$ at $J_y=0.4$ embedded in the integrability
  hyperboloid of $(J_y,J_z,A)$ space.}
\label{fig12}
\end{figure}

\begin{figure}[htb]
\centerline{\epsfig{file=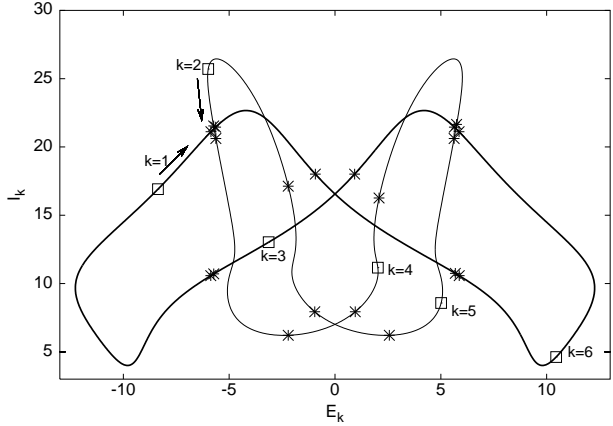,width=5.7cm,angle=-90}}
\caption{Open traces in the $(E_k,I_k)$-plane of all six $H_{A1A}^4$
  levels along the path with radius $\sqrt{A^2+J_z^2}=\sqrt{0.58}$ at $J_y=0.4$
  on the integrability hyperboloid in $(J_y,J_z,A)$ space. The traces start at
  the squares $(\alpha=0^\circ)$ in the directions indicated. The asterisks mark level
  crossing points.}
\label{fig13}
\end{figure}

The angular dependence of the six $H_{A1A}^4$ levels, depicted in Fig.
\ref{fig12}, does indeed not show any level collisions just as was the case in
Fig. \ref{fig2} for the first path. All levels undergo several crossings along
this path, and none of the crossings has any noticeable effect on the quantum
invariants $E_k, I_k$ plotted in Fig. \ref{fig13}. 

%%%%%%%%%%%%%%%%%BEGIN-FIGURE%%%%%%%%%%%%%%%%%%%%%
%%%%%%%%%%%%%%%%%%%%%END-FIGURE%%%%%%%%%%%%%%%%%%

%%%%%%%%%%%%%%%%%%%%BEGIN-FIGURE%%%%%%%%%%%%%%%%%%
%%%%%%%%%%%%%%%%%%%%%%END-FIGURE%%%%%%%%%%%%%%%%%

Nevertheless, there is a major difference between the evolution of eigenstates
along these two paths. Each one of the six levels shown in Fig.  \ref{fig12}
transforms into a different level in the course of one loop of the path around
the integrability hyperboloid. It takes three loops for every eigenstate to
return to its original position in the level sequence. On the plane of
invariants this phenomenon is reflected in open traces that connect to form two
rings of three segments each as shown in Fig. \ref{fig13}. The two sets of
levels are distinguished by line thickness.

Unlike in the previous situation (Sec. \ref{sec3C}), here the open-trace
phenomenon cannot be attributed to a change of symmetry along the path. What
distinguishes the first path, where open traces do not occur from the fourth
path, where they do occur, is that only the former can be shrunk to a point
without leaving the integrability manifold.  Hence the multiple connectedness of
the integrability hyperboloid forces us to allow for functions $H_Q(J_1,J_2)$
and $I_Q(J_1,J_2)$ whose dependence on the Hamiltonian parameters is still
smooth but multiple-valued.

With these concessions, the signature properties of quantum integrability
postulated above remain fully intact.  The quantum invariants $E_k, I_k$ exhibit
strongly contrasting features when observed along paths that are not embedded in
the integrability manifold. Visualizing these differences does not depend on a
statistical analysis. They are unmistakenly identifiable in systems for systems
with very few levels.

%%%%%%%%%%%%%%%%%%%%%%%%%%%%%%%%%%%%%%%%%%%%%
%
\subsection{Level repulsion due to nonintegrability}\label{sec3E}
%
%%%%%%%%%%%%%%%%%%%%%%%%%%%%%%%%%%%%%%%%%%%%%%

%%%%%%%%%%%%%%%%%%%BEGIN-FIGURE%%%%%%%%%%%%%%%%%%%%
\begin{figure}[t!]
\centerline{\epsfig{file=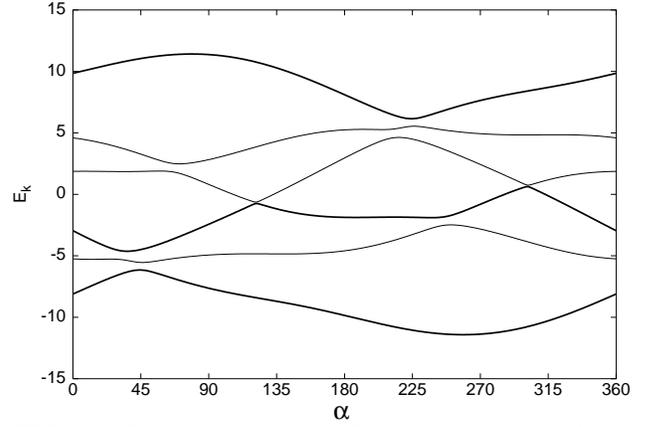,width=5.7cm,angle=-90}}
\caption{Energy eigenvalues $E_k, k=1,\ldots,6$ in the invariant subspace of
  $H_{A1A}^4$ plotted versus $\alpha$ on the path with radius
  $\sqrt{A^2+J_z^2}=\sqrt{0.3712}$ at $J_y=0.4$ off the integrability hyperboloid in
  $(J_y,J_z,A)$ space.}
\label{fig14}
\end{figure}
%%%%%%%%%%%%%%%%%%%%END-FIGURE%%%%%%%%%%%%%%%%%%%

%%%%%%%%%%%%%%%%%%%BEGIN-FIGURE%%%%%%%%%%%%%%%%%%%
\begin{figure}[t!]
\centerline{\epsfig{file=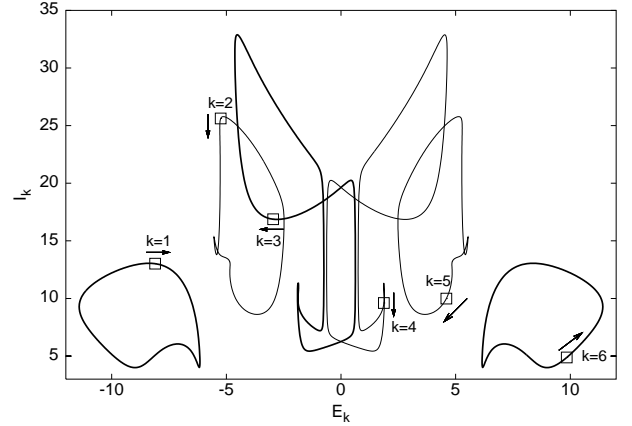,width=5.7cm,angle=-90}}
\caption{Closed traces in the $(E_k, I_k)$-plane of all six $H_{A1A}^4$
  levels along the path with radius $\sqrt{A^2+J_z^2}=\sqrt{0.3712}$ at
  $J_y=0.4$ off the integrability hyperboloid in $(J_y,J_z,A)$ space. The traces
  start at the the squares $(\alpha=0^\circ)$ in the directions indicated.}
\label{fig15}
\end{figure}
%%%%%%%%%%%%%%%%%%%%%END-FIGURE%%%%%%%%%%%%%%%%%%%

For a direct comparison with the previous situation, we now choose a circle with the same
center as the fourth path and a somewhat smaller radius, $J_z^2+A^2 =
0.3712$. This fifth path lies off the integrability manifold except for two
points where it intersects the integrability plane $A=0$ (see Fig. \ref{fig16}).
However, no level degeneracies occur at these intersection points.

The six $H_{A1A}^4$ levels versus $\alpha$ along the fifth path are plotted in Fig.
\ref{fig14}. Even though the resulting pattern is vaguely similar to that
observed in Fig. \ref{fig12}, the differences are clear-cut. All level crossings
have turned into level collisions.

Most of the collisions are fairly soft. The two hardest collisions are barely
resolved as such on the scale of Fig. \ref{fig15}. None of the levels transform
into each other any more. The levels are now naturally labeled by the energy
sorting quantum number. Each open segment of the traces shown in Fig.
\ref{fig13} has turned into a closed trace. All level collisions, especially the
hard ones, leave the characteristic marks on the traces in the form of a rapidly
varying second invariant $I_k$.

If we were to move the fifth path closer to the integrability hyperboloid by
increasing its radius (see Fig. \ref{fig16}), we could observe a gradual
hardening of all level collisions. The level configurations as shown in
Fig. \ref{fig14} would increasingly resemble those in Fig. \ref{fig12}. The
traces as shown in Fig. \ref{fig15}, however, would remain very different from
those pertaining to the integrable case (Fig. \ref{fig13}). 

Only in the limiting case where the fifth path merges with the fourth path would
the closed traces of the nonintegrable model break into segments connected by
vertical lines. The ends of each segment would then rejoin ends of other
segments to form the smooth rings of open traces shown in Fig. \ref{fig13}.

Similar observations are made upon lifting the first path off the integrability
plane $A=0$ to a plane at $A\neq 0$. All the level crossings that exist in Fig.
\ref{fig2}, for example, turn into level collisions. The closed traces such as
shown in Fig. \ref{fig3} break into pieces whose ends rejoin via near vertical
lines into a new set of closed traces.

Along the second path we had observed (in Fig. \ref{fig4}) level crossings (due
to integrability) and level collisions (due to nearby points of higher
symmetry). Lifting this path off the integrability plane again removes all level
crossings and results in a set of closed traces. The characteristic marks of
level collisions on the traces in the $(E_k, I_k)$-plane are the same no matter
whether they are caused by a reduced symmetry or by nonintegrability.

Lifting the third path off the integrability plane has the same effects on the
level crossings attributed to integrability and the level crossings attributed
to the higher symmetry at selected points in parameter space (Fig. \ref{fig6}).
All are removed indiscriminately.

%%%%%%%%%%%%%%%%%%%%%%%%%%%%%%%%%%%%%%%%%%%%%
%
\subsection{Open traces caused by nonintegrability}\label{sec3F}
%
%%%%%%%%%%%%%%%%%%%%%%%%%%%%%%%%%%%%%%%%%%%%%

%%%%%%%%%%%%%%%%%BEGIN-FIGURE%%%%%%%%%%%%%%%%%%%%%
\begin{figure}[t!]
\centerline{\epsfig{file=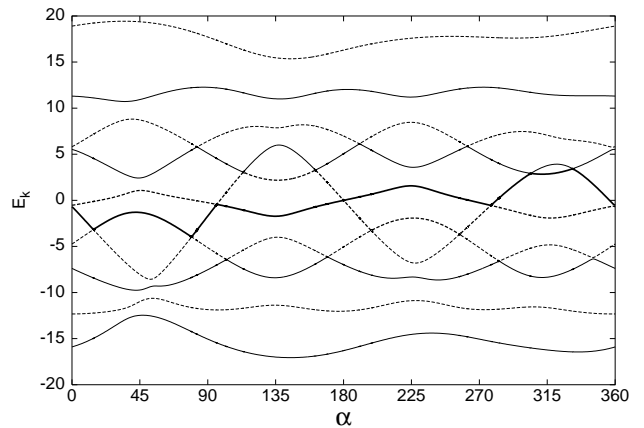,width=5.7cm,angle=-90}}
\caption{Energy eigenvalues $E_k, k=1,\ldots,10$ in the invariant subspace of
  $H_{A1A}^5$ plotted versus $\alpha$ on the path with radius
  $\sqrt{J_y^2+J_z^2}=0.5$ at $A=0.3\cos^2(\alpha/2)$ in $(J_y,J_z,A)$ space.}
\label{fig17}
\end{figure}
%%%%%%%%%%%%%%%%%%%%%END-FIGURE%%%%%%%%%%%%%%%%%%

%%%%%%%%%%%%%%%%%%%%BEGIN-FIGURE%%%%%%%%%%%%%%%%%%
\begin{figure}[t!]
\centerline{\epsfig{file=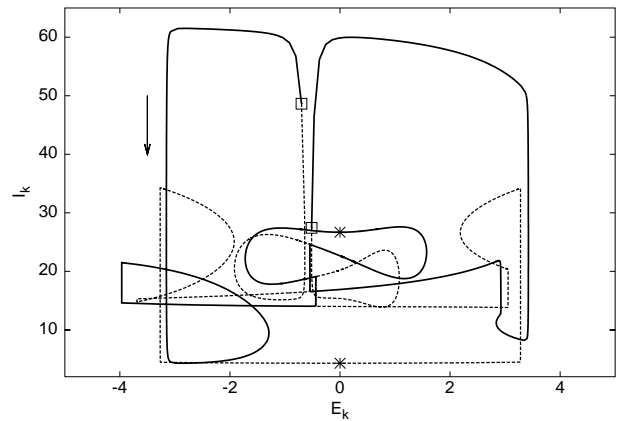,width=5.7cm,angle=-90}}
\caption{Open traces in the $(E_k, I_k)$-plane of the two $H_{A1A}^5$
  levels which undergoes one [5,6] crossings along the path with radius
  $\sqrt{J_y^2+J_z^2}=0.5$ at $A=0.3\cos^2(\alpha/2)$ in $(J_y,J_z,A)$ space. The
  traces start at the squares in the direction indicated. The asterisk on each
  trace marks the level crossing point at $\alpha=180^\circ$. }
\label{fig18}
\end{figure}
%%%%%%%%%%%%%%%%%%%%%END-FIGURE%%%%%%%%%%%%%%%%%%

The conflicting assignments of quantum numbers to eigenstates for parameter
values on and off the integrability manifold is most compellingly documented
when we pick a path in parameter space that is only partially embedded in the
integrability manifold.

The sixth path considered in this study of quantum invariants is a modification
of the first path (Sec. \ref{sec3A}) with the same projection in Fig.
\ref{fig1}. Whereas the first path was embedded in the integrability plane
$A=0$, the sixth path has a variable height relative to that plane: $A(\alpha)=
0.3\cos^2(\alpha/2)$. It touches down to the integrability plane at a single
point $(\alpha=180^\circ)$, where a [5,6] level crossing takes place.

Along this path there exist no other level crossings. All the other crossings
that existed in Fig. \ref{fig2} for the first path are now replaced by level
collisions (see Fig. \ref{fig17}).

The inevitable consequence of having a single level crossing along a closed path
in parameter space is the existence of a pair of open traces in the
plane of invariants, namely the traces of the states that undergo the [5,6] crossing
at $\alpha=180^\circ$. These traces are shown in Fig. \ref{fig18}. The ends of
the solid and dashed lines form a single loop, which is traced in the direction
indicated.

What causes here an open trace in the plane of invariants is obviously akin to
what had caused an open trace in the situation described in Sec. \ref{sec3C}. In
both cases two levels cross once due to particular circumstances at one point of
the path, and are thus prevented from crossing back to their original position
in the level sequence on the remaining stretch of the path. In Sec.~\ref{sec3C}
the particular circumstance was a higher symmetry, here it is integrability.

%%%%%%%%%%%%%%%%%%%%%%%%%%%%%%%%%%%%%%%%%%%%%
%
\section{Interpretation}\label{sec4}
%
%%%%%%%%%%%%%%%%%%%%%%%%%%%%%%%%%%%%%%%%%%%%%%

The study of quantum invariants along closed paths through parameter space
indicates that a change in symmetry and a change in integrability status produce
related phenomena. In some dynamical systems, the conservation laws that
guarantee integrability are direct consequences (via Noether's theorem) of
continuous symmetries. Switching from integrability to nonintegrability is then
accompanied by a reduction in symmetry.

In the two-spin model (\ref{Ham}), the presence of a (continuous rotational)
O(2) or higher symmetry in spin space does indeed imply the existence of a
second integral of the motion, namely the component of the total spin along the
symmetry axis, and integrability is guaranteed. However, a second integral of
the motion was shown to exist for certain parameter values even in the absence
of a continuous rotational symmetry.\cite{MTWKM87} Does integrability in that
case indicate the presence of a hidden symmetry? 

Classical integrability guarantees that the Hamiltonian (\ref{Ham}) can be
expressed as a function of the two action variables: $H=H_C(J_1,J_2)$. The
cyclical nature of the angle coordinates thus implies that $H_C$ is invariant
with respect to continuous rotation-like transformations in phase-space. Since
this is not related to a continuous symmetry in configuration space, it is
appropriate to call it a {\it hidden} symmetry.

For a description of the impact of symmetries on the level spectrum of the
quantum two-spin model, it is useful to distinguish three kinds of symmetry:
discrete symmetries, continuous symmetries, and hidden symmetries.

{\em Discrete} symmetries have no bearing on the classical integrability
property, but they do affect the shapes of phase-space trajectories. Quantum
mechanically, they divide the Hilbert space of $H$ into invariant subspaces.  In
general, this does not result in symmetry-induced level-degeneracies, but it
does lead to accidental degeneracies between levels belonging to different
invariant subspaces. Such level crossings exist independently of whether or not
$H$ is integrable.

{\em Hidden} symmetries, which guarantee classical integrability, cause additional
accidental level degeneracies, namely between states within one of the invariant 
subspaces pertaining to any existing discrete symmetry. 

{\em Continuous} symmetries, in essence, combine the effects of the discrete and
hidden symmetries, and allow accidental inter-subspace degeneracies. In addition
to these effects, continuous symmetries (sometimes in tandem with discrete
symmetries) produce level degeneracies of a permanent nature, the so-called
symmetry-induced level degeneracies.

There exists a hierarchy of symmetries in the two-spin model (\ref{Ham}): (S0)
In the absence of any symmetry, there are no level degeneracies. All levels will
collide when Hamiltonian parameters are varied. This situation can be realized
by an external magnetic field.  (S1) The existence of discrete symmetries alone
produces finite-D invariant Hilbert subspaces. Level crossings exist between
states belonging to different subspaces. Levels within any subspace collide.
(S2) The existence of hidden symmetries in addition to discrete symmetries
produces level crossings between states in the same invariant subspace. (S3) The
continuous symmetries produce permanent degeneracies in certain regions of
parameter space.

There exists a hierarchy of level collisions which corresponds to the hierarchy
of symmetries.  (S1$\to$S0) Inter-subspace level crossings in the presence of
discrete symmetries turn into level collisions when discrete symmetries are
removed.  (S2$\to$1) Intra-subspace level crossings turn into level collisions
when the hidden symmetries are removed, i.e. when the integrability is
destroyed.  (S3$\to$S2) Symmetry-induced level degeneracies associated with a
continuous symmetry are removed outside the range of that symmetry irrespective
of the presence or absence of the hidden symmetry.

Some level crossings along paths through symmetry points in parameter space turn
into level collisions along nearby paths that miss the symmetry point. Other
level crossings are insensitive to whether the path hits or misses the symmetry
point. They are the product of the hidden symmetry.

All phenomena observed in the quantum invariants $E_k,I_k$ along closed paths
on, off, and across the integrability manifold, indicate that the effects of a
change in integrability status are akin to the effects of a change in symmetry.
All observations point to the existence of a hidden symmetry that accompanies
quantum integrability.

In the classical limit, this hidden symmetry manifests itself in phase space
when viewed from a particular coordinate system -- the action-angle
coordinates. The same hidden symmetry must also exist in the quantum system, but
only on the integrability manifold. Even though nonintegrability is not to be
taken literally in the quantum case, the presence or absence of that hidden
symmetry has consequences that are equally clear-cut as in the classical limit.

\acknowledgments
This work was supported by the Research Office of the University of Rhode Island.
We are very grateful to Joachim Stolze for his comments and suggestions relating
to this work.

\begin{appendix}
%%%%%%%%%%%%%%%%%%%%%%%%%%%%%%%%%%%%%%%%%%%%%
%
\section{Discrete symmetries}\label{appA}
%
%%%%%%%%%%%%%%%%%%%%%%%%%%%%%%%%%%%%%%%%%%%%%
The (discrete) symmetry group relevant for the general 2-spin
Hamiltonian~(\ref{Ham}) is $D_2 \otimes S_2$, where $D_2$ contains the three
twofold rotations $C^\alpha_2$, $\alpha=x,y,z$ about the coordinate axes, and
$S_2$ the permutations of the two spins. The eight irreducible representations
of $D_2 \otimes S_2$ are named A1S, A1A, B1S, B1A, B2S, B2A, B3S, B3A,
where S (A) stand for (anti-)symmetric under permutation and A1, B1, B2, B3
for $(C_2^x, C_2^y, C_2^z) = (1,1,1)$, $(1,-1,-1)$, $(-1,1,-1)$,
$(-1,-1,1)$, respectively.\cite{SM90,Webe88}

%%%%%%%%%%%%%%%%%%%BEGIN-TABLE%%%%%%%%%%%%%%%%%%%
\begin{table}[hb]
\caption{Symmetry-adapted basis vectors for integer $\sigma$.
The local spin quantum numbers satisfy the relations $0<l_{1} \leq \sigma$,
$-l_{1}<l_{2}< l_{1}$. The subspace dimensionality $K$ is
$\frac{1}{2}(\sigma+1)(\sigma+2)$ for A1S, $\frac{1}{2}\sigma(\sigma-1)$ for
A1A, and $\frac{1}{2}\sigma(\sigma+1)$ for the other six classes.}
\label{tabeigv1}

\begin{tabular}{c|c}
A1S~~ & $\left| 00 \right>, \left( \left| l_{1},l_{1} \right> +
\left| -l_{1},-l_{1} \right> \right)/ \sqrt{2}$, \\
& $\left( \left| l_{1},-l_{1} \right> +\left| -l_{1},l_{1}\right>
\right)/ \sqrt{2}$,  \\
& $\left( \left| l_{1},l_{2} \right> +\left| -l_{1},-l_{2} \right> +
 \left| l_{2},l_{1} \right> +\left| -l_{2},-l_{1} \right> \right)/ 2$ \\
& $l_{1}+l_{2}$ even. \\ \hline
A1A~~ & 
$\left( \left| l_{1},l_{2} \right> +\left| -l_{1},-l_{2} \right> -
 \left| l_{2},l_{1} \right> -\left| -l_{2},-l_{1} \right> \right)/ 2$ \\
 & $l_{1}+l_{2}$ even. \\ \hline
B1S~~ & 
$\left( \left| l_{1},l_{1} \right> -\left| -l_{1},-l_{1} \right>
\right) / \sqrt{2}$ \\
& $\left( \left| l_{1},l_{2} \right> -\left| -l_{1},-l_{2} \right> +
 \left| l_{2},l_{1} \right> -\left| -l_{2},-l_{1} \right> \right)/ 2$ \\
& $l_{1}+l_{2}$ even. \\ \hline
B1A~~ & 
$\left( \left| l_{1},-l_{1} \right> -\left| -l_{1},l_{1} \right>
\right) / \sqrt{2}$ \\
& $\left( \left| l_{1},l_{2} \right> -\left| -l_{1},-l_{2} \right> -
 \left| l_{2},l_{1} \right> +\left| -l_{2},-l_{1} \right> \right)/ 2$ \\
& $l_{1}+l_{2}$ even. \\ \hline
B2S~~ & 
$\left( \left| l_{1},l_{2} \right> -\left| -l_{1},-l_{2} \right> +
 \left| l_{2},l_{1} \right> -\left| -l_{2},-l_{1} \right> \right)/ 2$ \\
& $l_{1}+l_{2}$ odd. \\ \hline
B2A~~ & 
$\left( \left| l_{1},l_{2} \right> -\left| -l_{1},-l_{2} \right> -
 \left| l_{2},l_{1} \right> +\left| -l_{2},-l_{1} \right> \right)/ 2$ \\
& $l_{1}+l_{2}$ odd. \\ \hline
B3S~~ & 
$\left( \left| l_{1},l_{2} \right> +\left| -l_{1},-l_{2} \right> +
 \left| l_{2},l_{1} \right> +\left| -l_{2},-l_{1} \right> \right)/ 2$ \\
& $l_{1}+l_{2}$ odd. \\ \hline
B3A~~ & 
$\left( \left| l_{1},l_{2} \right> +\left| -l_{1},-l_{2} \right> -
 \left| l_{2},l_{1} \right> -\left| -l_{2},-l_{1} \right> \right)/ 2$ \\
& $l_{1}+l_{2}$ odd. \\ 
\end{tabular}
\end{table}
%%%%%%%%%%%%%%%%%%%%END-TABLE%%%%%%%%%%%%%%%%%%%%

%%%%%%%%%%%%%%%%%%%%BEGIN-TABLE%%%%%%%%%%%%%%%%%%%
\begin{table}[hbt]
\caption{Symmetry-adapted basis vectors for half-integer $\sigma$.
The local spin quantum numbers satisfy the 
relations $0<l_{1} \leq \sigma$,
$-l_{1}<l_{2}< l_{1}$. The subspace dimensionality $K$ is
$\frac{1}{8}(4\sigma^2-1)$ for symmetric and
  $\frac{1}{8}(2\sigma+1)(2\sigma+3)$ for antisymmetric representations.}
\label{tabeigv2}
\begin{tabular}{c|c}
A1S~~ &
$\left( \left| l_{1},l_{2} \right> -\left| -l_{1},-l_{2} \right> +
 \left| l_{2},l_{1} \right> -\left| -l_{2},-l_{1} \right> \right)/ 2$ \\
& $l_{1}+l_{2}$ even. \\ \hline
A1A~~ & 
$\left( \left| l_{1},-l_{1} \right> -\left| -l_{1},l_{1} \right>\right)
/ \sqrt{2}$ \\ 
& $\left( \left| l_{1},l_{2} \right> -\left| -l_{1},-l_{2} \right> -
 \left| l_{2},l_{1} \right> +\left| -l_{2},-l_{1} \right> \right)/ 2$ \\
& $l_{1}+l_{2}$ even. \\ \hline
B1S~~ & 
$\left( \left| l_{1},-l_{1} \right> +\left| -l_{1},l_{1} \right>
\right)/ \sqrt{2}$ \\ 
& $\left( \left| l_{1},l_{2} \right> +\left| -l_{1},-l_{2} \right> +
 \left| l_{2},l_{1} \right> +\left| -l_{2},-l_{1} \right> \right)/ 2$ \\
& $l_{1}+l_{2}$ even. \\ \hline
B1A~~ & 
$\left( \left| l_{1},l_{2} \right> +\left| -l_{1},-l_{2} \right> -
 \left| l_{2},l_{1} \right> -\left| -l_{2},-l_{1} \right> \right)/ 2$ \\
& $l_{1}+l_{2}$ even. \\ \hline
B2S~~ & 
$\left( \left| l_{1},l_{1} \right> +\left| -l_{1},-l_{1} \right>
\right) / \sqrt{2}$ \\ 
& $\left( \left| l_{1},l_{2} \right> +\left| -l_{1},-l_{2} \right> +
 \left| l_{2},l_{1} \right> +\left| -l_{2},-l_{1} \right> \right)/ 2$ \\
& $l_{1}+l_{2}$ odd. \\ \hline
B2A~~ & 
$\left( \left| l_{1},l_{2} \right> +\left| -l_{1},-l_{2} \right> -
 \left| l_{2},l_{1} \right> -\left| -l_{2},-l_{1} \right> \right)/ 2$ \\
& $l_{1}+l_{2}$ odd. \\ \hline
B3S~~ & 
$\left( \left| l_{1},l_{1} \right> -\left| -l_{1},-l_{1} \right>
\right) / \sqrt{2}$ \\ 
& $\left( \left| l_{1},l_{2} \right> -\left| -l_{1},-l_{2} \right> +
 \left| l_{2},l_{1} \right> -\left| -l_{2},-l_{1} \right> \right)/ 2$ \\
& $l_{1}+l_{2}$ odd. \\ \hline
B3A~~ & 
$\left( \left| l_{1},l_{2} \right> -\left| -l_{1},-l_{2} \right> -
 \left| l_{2},l_{1} \right> +\left| -l_{2},-l_{1} \right> \right)/ 2$ \\
& $l_{1}+l_{2}$ odd. \\ 
\end{tabular}
\end{table}
%%%%%%%%%%%%%%%%%%%%%%END-TABLE%%%%%%%%%%%%%%%%%%

The basis vectors with transformation properties corresponding to the eight
different irreducible representations $R$ are listed in Table~\ref{tabeigv1} for
integer $\sigma$ and in Table~\ref{tabeigv2} for half-integer $\sigma$. The
Hamiltonian matrix can then be expressed in the form
\begin{equation}\label{HRsigma}
H = \bigoplus_{R,\sigma} H_R^\sigma
\end{equation}
with blocks of dimensionalities $K = 1, 3, 6, 10, \ldots$ in 16 different
realizations, two for each symmetry class (one with integer $\sigma$ and one
with half-integer $\sigma$).\cite{note1}

\end{appendix}

%%%%%%%%%%%%%%%%%%%%%%%%%%%%%%%%%%%%%%%%%%%%%%
%


\begin{thebibliography}{99}
\bibitem{note4}
Only one exception to that rule is known. The energy level $E=0$ is also
degenerate on a 3D manifold away from the integrability hypersurface (\ref{Int}).
(see Refs. \onlinecite{SM98}, \onlinecite{Webe88}).
\bibitem{note3} A natural choice of $A$ would be an operator that commutes with
  $H$ on the integrability hypersurface as can be constructed from the known
  second classical integral of the motion (Ref.~\onlinecite{MTWKM87}), but any
  polynomial of the $S_i^\alpha$ will produce expectation values with equivalent
  properties (Refs.~\onlinecite{Pere84,MTWKM87}). Only the diagonal matrix
  elements matter in this context.
\bibitem{note5} The $J_\alpha, A_\alpha$ are measured in arbitrary energy units divided
  by $\hbar^2$. Henceforth, this unit will be suppressed in all explicit results
  that follow.
\bibitem{note2} The levels involved in the crossings are $[k,k+1]$ with
  $k=5,7,4,4,7,5,3,6,6,3,5$ starting from either end of the horizontal line
  counterclockwise along the frame of the graph.
\end{thebibliography}
\end{document}